\documentclass[11pt,a4paper]{article}       % onecolumn (second format)
\usepackage{graphicx}
\usepackage[misc,geometry]{ifsym}
\usepackage{amsmath}
\usepackage{amsfonts}
\usepackage{amssymb}
\usepackage[american]{babel}
\usepackage{tikz}
\usepackage{subfigure}
\usepackage{url}
\newcommand{\ket}[1]{\big| #1 \big\rangle}
\newcommand{\bra}[1]{\big\langle #1 \big|}
\newcommand{\braket}[2]{\big\langle #1 \big| #2 \big\rangle}                 % < | >
\newcommand{\bracket}[3]{\big\langle #1 \big| #2 \big| #3 \big\rangle}       % < | | >
\newcommand{\abs}[1]{\left| #1 \right|}

\newcommand{\vk}{\vec{k}}
\newcommand{\vzero}{\vec{0}}
\newcommand{\vm}{\vec{m}}

\newcommand{\vbeta}{\vec{\beta}}

\newcommand{\mybar}[1]{#1^*}
\newcommand{\grafico}{\begin{tikzpicture}[baseline=6.3ex,scale=1]
    \draw [<->,thick] (0,2.5) node (yaxis) [left] {\scriptsize $\beta$}
        |- (2.5,0) node (xaxis) [below] {\scriptsize $\alpha$};
    \draw (0,2) node[left] {\scriptsize $\pi$} -- (2,2) ;
    \draw (2,0) node[below] {\scriptsize $\pi$} -- (2,2) ;
    \draw (0,0) node[left] {\scriptsize $0$} -- (2,2) ;
    \draw (0,2)  -- (2,0) ;
    \draw (1,1.5) node {\scriptsize II}; %label
    \draw (1,0.5) node {\scriptsize I}; %label
    \draw (0.5,1) node {\scriptsize III}; %label
     \draw (1.5,1) node {\scriptsize IV}; %label
     \end{tikzpicture}}

\begin{document}

\title{Moments of Coinless Quantum Walks on Lattices%\thanks{Grants or other notes
}


\author{Raqueline A. M. Santos$^1$,  Renato Portugal$^{1,2}$, Stefan Boettcher$^3$\\
\\
{\small $^1$Laborat\'{o}rio Nacional de Computa\c{c}\~{a}o Cient\'{i}fica}\\
{\small Petr\'{o}polis, RJ 25651-075, Brazil}\\
{\small raqueline@lncc.br, portugal@lncc.br}\\
\\
{\small $^2$Universidade Cat\'olica de Petr\' opolis}\\
{\small Petr\'{o}polis, RJ, 25685-070,  Brazil}\\\\
{\small $^3$Department of Physics, Emory University}\\
{\small Atlanta, GA 30322, USA}\\
{\small sboettc@emory.edu }
}
\date{}

\maketitle

\begin{abstract}
The properties of the coinless quantum walk model have not been as thoroughly analyzed as those of the coined model.  Both evolve in discrete time steps but the former uses a smaller Hilbert space, which is spanned merely by the site basis. Besides, the evolution operator can be obtained using a process of lattice tessellation, which is very appealing. The moments of the probability distribution play an important role in the context of quantum walks. The ballistic behavior of the mean square displacement indicates that quantum-walk-based algorithms are faster than random-walk-based ones. In this paper,
we obtain analytical expressions for the moments of the coinless model on  $d$-dimensional lattices. The mean square displacement for large times is explicitly calculated for the one- and two-dimensional lattices and, using optimization methods, the parameter values that give the largest spread are calculated and compared with the equivalent ones of the coined model. 
Although we have employed asymptotic methods, 
our approximations are accurate even for small numbers of time steps.
%\keywords{coinless quantum walks \and moments \and mean square displacement \and standard deviation}
\end{abstract}

\section{Introduction}
 
Quantum walks are the quantum versions of random walks. Their interesting non-classical behavior has allowed the development of faster quantum search algorithms~\cite{Portugal:2013}. Discrete-time quantum walks, introduced by Aharonov {\it et al}.~\cite{Aharonov:1993},  use an additional space which represents the coin, whereas coinless quantum walks, introduced by Patel {\it et al.}~\cite{Patel:2005}, have a Hilbert space which is spanned only by the site basis. The coined model has been widely studied during the last decade. For example, Ambainis {\it et al}.~\cite{Ambainis:2001} analyzed the dynamics of one-dimensional coined walks and showed that they spread quadratically faster compared to the classical random walks. Moments and the mean square displacement (or variance) of the coined quantum walk on one-dimensional lattices were analyzed in Refs.~\cite{Nayak:2000,Konno:2002,SBJ14,FB14} and on two-dimensional lattices in Refs.~\cite{WKKK08,Ampadu:2011,PBHMPBK14}.

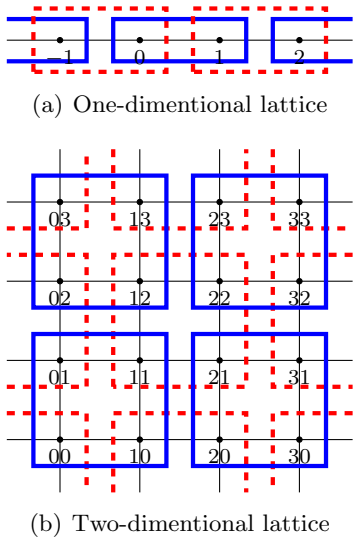
\begin{figure}[!htb]
\centering
\subfigure[fig1][One-dimentional lattice]{
\begin{tikzpicture}[scale=0.35]
\draw (-1,0) -- (12,0);
\draw [blue,ultra thick] (-1,0.8) -- (2,0.8) -- (2,-0.8) -- (-1,-0.8);
\draw [red,dashed,ultra thick] (0,-1.2) rectangle (5,1.2);
\draw [blue,ultra thick] (3,-0.8) rectangle (8,0.8);
\draw [red,dashed,ultra thick] (6,-1.2) rectangle (11,1.2);
\draw [blue,ultra thick] (12,0.8) -- (9,0.8) -- (9,-0.8) -- (12,-0.8);
\draw[fill] (1,0) circle [radius=0.1] node[below] {\scriptsize{$-1$}};
\draw[fill] (4,0) circle [radius=0.1] node[below] {\scriptsize{$0$}};
\draw[fill] (7,0) circle [radius=0.1] node[below] {\scriptsize{$1$}};
\draw[fill] (10,0) circle [radius=0.1] node[below] {\scriptsize{$2$}};
\end{tikzpicture}  }
\\
\subfigure[fig2][Two-dimentional lattice]{
\begin{tikzpicture}[scale=0.35]
\draw (-1,0) -- (12,0);
\draw (-1,3) -- (12,3);
\draw (-1,6) -- (12,6);
\draw (-1,9) -- (12,9);
\draw (1,-2) -- (1,11);
\draw (4,-2) -- (4,11);
\draw (7,-2) -- (7,11);
\draw (10,-2) -- (10,11);
\draw [dashed,red,ultra thick] (-1,1) -- (2,1) -- (2,-2);
\draw [dashed,red,ultra thick] (-1,2) -- (2,2) -- (2,7) -- (-1,7);
\draw [dashed,red,ultra thick] (3,11) -- (3,8) -- (8,8) -- (8,11);
\draw [dashed,red,ultra thick] (-1,8) -- (2,8) -- (2,11);
\draw [dashed,red,ultra thick] (9,11) -- (9,8) -- (12,8);
\draw [dashed,red,ultra thick] (12,7) -- (9,7) -- (9,2) -- (12,2);
\draw [dashed,red,ultra thick] (3,-2) -- (3,1) -- (8,1) -- (8,-2);
\draw [dashed,red,ultra thick] (12,1) -- (9,1) -- (9,-2);
\draw [dashed,red,ultra thick] (3,2) rectangle (8,7);
\draw [blue,ultra thick] (0,-1) rectangle (5,4);
\draw [blue,ultra thick] (6,-1) rectangle (11,4);
\draw [blue,ultra thick] (6,5) rectangle (11,10);
\draw [blue,ultra thick] (0,5) rectangle (5,10);
\draw[fill] (1,0) circle [radius=0.1] node[below] {\scriptsize{$00$}};
\draw[fill] (4,0) circle [radius=0.1] node[below] {\scriptsize{$10$}};
\draw[fill] (7,0) circle [radius=0.1] node[below] {\scriptsize{$20$}};
\draw[fill] (10,0) circle [radius=0.1] node[below] {\scriptsize{$30$}};
\draw[fill] (1,3) circle [radius=0.1] node[below] {\scriptsize{$01$}};
\draw[fill] (4,3) circle [radius=0.1] node[below] {\scriptsize{$11$}};
\draw[fill] (7,3) circle [radius=0.1] node[below] {\scriptsize{$21$}};
\draw[fill] (10,3) circle [radius=0.1] node[below] {\scriptsize{$31$}};
\draw[fill] (1,6) circle [radius=0.1] node[below] {\scriptsize{$02$}};
\draw[fill] (4,6) circle [radius=0.1] node[below] {\scriptsize{$12$}};
\draw[fill] (7,6) circle [radius=0.1] node[below] {\scriptsize{$22$}};
\draw[fill] (10,6) circle [radius=0.1] node[below] {\scriptsize{$32$}};
\draw[fill] (1,9) circle [radius=0.1] node[below] {\scriptsize{$03$}};
\draw[fill] (4,9) circle [radius=0.1] node[below] {\scriptsize{$13$}};
\draw[fill] (7,9) circle [radius=0.1] node[below] {\scriptsize{$23$}};
\draw[fill] (10,9) circle [radius=0.1] node[below] {\scriptsize{$33$}};
\end{tikzpicture}  }
\caption{Example of tessellations for the one- and two-dimensional lattices. $U_0$ is associated to the blue tessellation (solid line) and $U_1$ is associated to the red tessellation (dashed line). }
\label{fig:tess}
\end{figure}

 The coinless or staggered quantum walk model is defined by an evolution operator that is the product of two reflections, $U_0$ and $U_1$, acting on the site basis. These reflections can be obtained through a process of lattice tessellation as described by Falk~\cite{Falk:2013}. Examples of tessellations for one- and two-dimensional lattices are depicted in Fig.~\ref{fig:tess}.
Different tessellations can be used to the generic $d$-dimensional lattice but some of them generate operators which describe trivial walks.  Quantum search algorithms on two-dimensional lattices using the coinless model were analyzed numerically in Refs.~\cite{Patel:2005,Patel:2010,Falk:2013}. The numerical results suggested that this model is as efficient as the coined model with the advantage of using a smaller Hilbert space.  Ambainis \textit{et. al}~\cite{Ambainis:2013} proved analytically that the coinless model finds a marked site in time $O(\sqrt{N\log N})$ for a two-dimensional lattice with $N$ vertices using Falk's model, confirming the numerical results just mentioned.  Portugal {\it et al}.~\cite{Portugal:2014} analyzed the dynamics of one-dimensional coinless walks and its relation with the coined model.

The coinless model has not been so extensively analyzed as the coined model. Specially important in this context are the moments of the probability distribution. The mean square displacement, for example, gives us information about how far from the initial position a walker can be found. If quantum walks spread faster than random walks, there is hope for improving random-walk-based algorithms by using quantum walks.
In this paper, we analyze the moments of the coinless model on lattices. Due to the translational invariance, it is possible to find a Fourier transform that generates a $2d\times 2d$ reduced evolution operator, which contains all information about the dynamics.  After calculating the eigenvalues of  this reduced operator, we obtain an analytical expression of the $n$th moment in terms of the $n$th derivative of  the eigenvalues and give explicit solutions for the one- and two-dimensional lattices. For the one-dimensional lattice we use the most generic coinless quantum walk with a 2-site tessellation, see  Fig.~\ref{fig:tess}, taking a localized initial condition. For the two-dimensional lattice we use a 4-site tessellation taking the simplest basis vectors with non-localized initial conditions. For both cases, we analyze the mean square displacement and obtain what are the best choice for the largest spread and compare with the results of the coined model.

This paper is organized as follows. In Sec.~\ref{sec:qw} we describe the coinless model on lattices. In Sec.~\ref{sec:m}, we obtain an analytical expression for the moments of the coinless model using the moment generating function. The moments and the mean square displacement are calculated for the one- and two-dimensional lattices in Sec.~\ref{sec:1d} and Sec.~\ref{sec:2d}, respectively. In the Appendix, the first and second moments of the coined model on one- and two-dimensional lattices are given for comparison. 

\section{Coinless quantum walks on lattices}\label{sec:qw}
The equation that describes the evolution of the quantum walk is 
\begin{equation}
\ket{\psi(t)} = {\cal{U}}^t\ket{\psi(0)},
\label{propsolution}
\end{equation}
where ${\cal{U}}=U_{1} U_{0}$ is the propagator. The reflections $U_0$ and $U_1$ are defined as
\begin{equation}
U_{0,1} = 2\sum_{\vec{x}}\ket{u_{\vec{x}}^{0,1}}\bra{u_{\vec{x}}^{0,1}} - \cal{I},
\end{equation}
%and the sum runs over all lattice sites $\vec{x}$.
where the sum runs over all patches of the associated tessellation.
As shown by Fig.~\ref{fig:tess}, we have two different tessellations which generate reflections $U_{0}$ and $U_{1}$. Vectors $\ket{u_{\vec{x}}^{0,1}}$ are superpositions of $2d$ vertices associated with one patch of the tessellation, and can be written as
\begin{eqnarray}
\ket{u_{\vec{x}}^{0}} &=& \sum_{\vbeta \in \{0,1\}^{d}}u_{\vbeta}^{0}\ket{2\vec{x}+\vbeta}, \\
\ket{u_{\vec{x}}^{1}} &=& \sum_{\vbeta \in \{0,1\}^{d}}u_{\vbeta}^{1}\ket{2\vec{x}+\vbeta+1}, 
\end{eqnarray}
with $\sum_{\vbeta \in \{0,1\}^{d}}\left| u_{\vbeta}^{0,1} \right|^2 = 1$.
%\subsection{Fourier solution}\label{sec:fsol}

The eigenspectrum of ${\cal{U}}$ is obtained using a staggered Fourier transform, which is defined by
\begin{equation}\label{eq:fbasis}
\ket{\psi_{\vk}^{ \vbeta }} = \sum_{\vm \textrm{ even}}e^{-i(\vm+\vbeta)\cdot\vk}\ket{\vm+\vbeta},
\end{equation}
where $\vbeta \in \{0,1\}^{d}$ and $\vm$ runs over sites with even labels. 
%represents the sites of the form $(2r,2s)$, with $r,s \in (-\infty,\infty)$.
The staggered Fourier basis spans a hyperplane, which is invariant under the action of ${\cal{U}}$. A vector in this hyperplane can be represented by a reduced vector with $2d$ entries. The action of  ${\cal{U}}$ on the reduced vector can be described by a reduced operator ${{\cal{U}}_{\vk}}$, which depends on $\vk$. Let $\ket{v_{\vk}^{ \vbeta }}$ and $\ket{w_{\vk}^{ \vbeta }}$ be the eigenvectors of ${\cal{U}}$ and ${{\cal{U}}_{\vk}}$, respectively. 
The relation that connects these eigenvectors is
\begin{equation}\label{eq:vk}
\ket{v_{\vk}^{ \vbeta }} = \sum_{\vec{\beta^{\prime}}\in \{0,1\}^{d}}\braket{\vec{\beta^{\prime}}}{w_{\vk}^{ \vbeta }}\ket{\psi_{\vk}^{ \vec{\beta^{\prime}} }}.
\end{equation}
The eigenvalues $\lambda_{\vk}^{ \vbeta }$ of ${\cal{U}}$ and ${{\cal{U}}_{\vk}}$ are the same.
For $\vec{k} = \{k_1,\dots,k_d\}$, let us denote
\begin{equation*}
\int_{-\pi}^{\pi}\cdots \int_{-\pi}^{\pi}dk_1\dots dk_d\quad\textrm{ by }\quad\int_{-\pi}^{\pi}d\vk.
\end{equation*}
The connection between ${\cal{U}}$ and ${\cal{U}}_{\vk}$ is 
\begin{equation}\label{eq:U2t}
{\cal{U}}  =  \int_{-\pi}^{\pi}\frac{d\vk}{(2\pi)^d}\sum_{\vec{\beta},\vec{\beta^{\prime}}}\bra{\vec{\beta}}{{\cal{U}}_{\vk}}\ket{\vec{\beta^{\prime}}}\ket{\psi_{\vk}^{ \vec{\beta} }}\bra{\psi_{\vk}^{ \vec{\beta^{\prime}} }}.
\end{equation}
The $t$-th power ${\cal{U}}^t$ is obtained by substituting ${{\cal{U}}_{\vk}}$ for ${\cal{U}}_{\vk}^t$, which follows from repeated insertions of the completeness relation in the Fourier basis and $\braket{\psi_{\vk}^{ \vec{\beta^{\prime}} }}{\psi_{\vk}^{ \vec{\beta} }}=\delta_{\vec{\beta},\vec{\beta^{\prime}}}\delta_{\vec{k},\vec{k^\prime}}$.
%One of the interesting features of the quantum walk is that it exhibits a quadratically faster rate of spread (measured using variance) compared to the classical random walk. The variance is defined as, ?2(t) = ?xpx(t) á (x ? µ(t))2, µ(t) = ?xpx(t) á x, where px(t) is the total probability of finding the walker at location x at time t, given by summing over the coin degree of freedom, px(t) = ?c|?x, c(t)|2, effectively tracing out the coin.

\section{Moments}\label{sec:m}
Let us calculate the moment generating function $\left< e^{ik_jx_j}\right>_t = \bracket{\psi(t)}{e^{ik_jx_j}}{\psi(t)}$ at time $t$. Recall that $\vec{k} = \{k_1,\dots,k_d\}$ and $\vec{x} = \{x_1,\dots,x_d\}$. Using  Eqs.~(\ref{propsolution}) and~(\ref{eq:U2t}), we eventually obtain
\begin{equation}\label{eq:expikx_t}
\left< e^{ik_jx_j}\right>_t = \int_{-\pi}^{\pi}\frac{d\vec{k^\prime}}{(2\pi)^d}\sum_{\vbeta,\vec{\beta^{\prime}}}\bra{\vbeta}({{\cal{U}}_{\vec{k^{\prime}}}^t})^\dagger {\cal{U}}_{\vec{k^{\prime\prime}}}^t \ket{\vec{\beta^\prime}}\braket{\psi(0)}{\psi_{\vec{k}^{\prime}}^{{\vbeta}}}\braket{\psi_{\vec{k}^{\prime\prime}}^{{\vbeta}^{\prime}}}{\psi(0)},
\end{equation}
where $\vec{k}^{\prime\prime}=\{k'_1,\dots,k'_{j-1},k'_j+k_j,\dots,k'_d\}$.
Let us take the initial condition $\ket{\psi(0)} = \ket{\vzero}$. Then $\braket{\psi(0)}{\psi_{\vec{k}}^{\vec{\beta}}}=\delta_{\vbeta,\vzero}$ for any $\vk$ and
\begin{equation}
\left< e^{ik_jx_j}\right>_t = \int_{-\pi}^{\pi}\frac{d\vec{k^\prime}}{(2\pi)^d}\bra{\vzero}({{\cal{U}}_{\vec{k^{\prime}}}^t})^\dagger {\cal{U}}_{\vec{k^{\prime\prime}}}^t\ket{\vzero}.
\end{equation}
By differentiating the moment generating function $n$ times with respect to $k_j$ and setting $k_j = 0$ we obtain the $n$-th moment at time $t$
\begin{eqnarray}\
\left< x_j^n\right>_t &=& \left(-i\frac{\delta}{\delta k_j}\right)^n\left< e^{ik_jx_j}\right>_t\Big|_{k_j = 0}\\
&=& \int_{-\pi}^{\pi}\frac{d\vec{k}}{(2\pi)^d}\bra{\vzero}({{\cal{U}}_{\vec{k}}^t})^\dagger \left[\left(-i\frac{\delta}{\delta k_j}\right)^n{\cal{U}}_{\vec{k}}^t\right]\ket{\vzero}.\label{eq:moment}
\end{eqnarray}

Let 
\begin{equation}
\Lambda_{\vk} = \left[\ket{w_{\vk}^{ \vbeta }} \right]_{\vbeta}
\end{equation}
be the diagonalizing matrix, whose columns are the eigenvectors of ${\cal{U}}_{\vk}$, such that ${{\cal{U}}_{\vk}} = \Lambda_{\vk} D\left[\lambda^{ \vbeta}_{\vk}\right]\Lambda_{\vk}^\dagger$, where we define $D$ as the diagonal matrix of the eigenvalues.
Then,
\begin{equation}\label{eq:diag}
{{\cal{U}}_{\vk}}^t = \Lambda_{\vk} D\left[\left(\lambda^{ \vbeta}_{\vk}\right)^t\right]\Lambda_{\vk}^\dagger,
\end{equation}
after repeated use of  $\Lambda_{\vk}^\dagger \Lambda_{\vk} = I$.
The right hand side of Eq.~(\ref{eq:diag}) oscillates with respect to $t$ because the eigenvalues of ${\cal{U}}_{\vk}$ have modulus $1$. The $n$-th derivative with respect to $k_j$ generates terms proportional to $t^n$.
Using Eq.~(\ref{eq:diag}) we obtain to leading order in $t$:
\begin{eqnarray*}
\left(-i\frac{\delta}{\delta k_j}\right)^n{{\cal{U}}_{\vk}}^t &=& t^n\Lambda_{\vk} D\left[\left(\lambda^{ \vbeta}_{\vk}\right)^t\left(-i\frac{\delta \ln\lambda^{ \vbeta }_{\vk} }{\delta k_j}\right)^n\right]\Lambda_{\vk}^\dagger+O(t^{n-1}),
\end{eqnarray*}
and after substituting into Eq.~(\ref{eq:moment}), we obtain
\begin{eqnarray}\label{eq:xn}
\left<x_j^n\right>_t &=&  t^n\int_{-\pi}^{\pi}\frac{d\vk}{(2\pi)^d}\bra{\vzero}\Lambda_{\vk} D\left[\left(-i\frac{\delta \ln\lambda^{ \vbeta }_{\vk} }{\delta k_j}\right)^n\right]\Lambda_{\vk}^\dagger\ket{\vzero}+O(t^{n-1}), \label{eq:xntf}
\end{eqnarray}
again using  $\Lambda_{\vk}^\dagger \Lambda_{\vk} = I$ repeatedly, and realizing that both diagonal matrices merge into one matrix using that $\left|\lambda^{ \vbeta }_{\vk}\right|^2=1$.

\section{One-dimensional lattice}\label{sec:1d}
From the tessellation for the one-dimensional lattice depicted in Fig.~\ref{fig:tess}, we define the vectors
\begin{eqnarray}
\ket{u^0_x} &=& \cos\frac{\alpha}{2}\ket{2x} + e^{i\phi_1}\sin\frac{\alpha}{2}\ket{2x+1},\\
\ket{u^1_x} &=& \cos\frac{\beta}{2}\ket{2x+1} + e^{i\phi_2}\sin\frac{\beta}{2}\ket{2x+2},
\end{eqnarray}
with tunable parameters $(\alpha, \phi_1)$ and $(\beta, \phi_2)$, each pair generating a Bloch sphere. The propagator is ${\cal{U}} = U_1U_0$, where
\begin{equation}
U_{0,1} = 2\sum_{x=-\infty}^{\infty}\ket{u_x^{0,1}}\bra{u_x^{0,1}}-\cal{I}.
\end{equation}
Ref.~\cite{Portugal:2014} describes the eigenvalues and eigenvectors for the coinless quantum walk on the one-dimensional lattice. The Fourier basis in this case is given by
\begin{eqnarray}
\ket{\psi_k^{0}} &=& \sum_{x=-\infty}^\infty e^{-2xki}\ket{2x},\\
\ket{\psi_k^{1}} &=& \sum_{x=-\infty}^\infty e^{-(2x+1)ki}\ket{2x+1}.
\end{eqnarray}
The reduced matrix in this case is
\begin{equation}
{\cal{U}}_{k} = \left(\begin{array}{cc}A & -B^* \\B & A^*\end{array}\right),
\end{equation}
where
\begin{eqnarray}
A&=&-\cos\alpha\cos\beta+\sin\alpha\sin\beta e^{i(2k+\phi_1+\phi_2)},\\
B&=&\sin\alpha\cos\beta e^{i(k+\phi_1)}+\cos\alpha\sin\beta e^{-i(k+\phi_2)}.
\end{eqnarray}
The eigenvalues of ${{\cal{U}}_{k}}$ are $\lambda = e^{\pm i\theta}$, where
\begin{equation}
\cos\theta = \frac{A+A^*}{2}.
\end{equation}
The associated eigenvectors are
\begin{equation}\label{eq:vec}
\frac{1}{\sqrt{C^{\pm}}}\left(\begin{array}{c}-B^* \\e^{\pm i\theta} - A\end{array}\right),
\end{equation}
where
\begin{equation}
C^\pm = \sin\theta(2\sin\theta \pm i(A-A^*)).
\end{equation}

Using Eq.~(\ref{eq:vec}) and
\begin{equation}
\frac{\delta \theta}{\delta k} = \frac{A-A^*}{i\sin\theta},
\end{equation}
we obtain from Eq.~(\ref{eq:xntf}) that the odd moments are
\begin{equation}\label{eq:oddm}
\left< x^{2n-1}\right>_t = \frac{t^{2n-1}}{4\pi}\int_{-\pi}^{\pi}\left[\frac{A-A^*}{i\sin\theta}\right]^{2n}dk+O(t^{2n-2}),
\end{equation}
and the even moments are
\begin{equation}\label{eq:evenm}
\left< x^{2n}\right>_t = 2t\left< x^{2n-1}\right>_t +O(t^{2n-1}).
\end{equation}

\begin{figure}[!htb]
\centering
\includegraphics[width=3.5in]{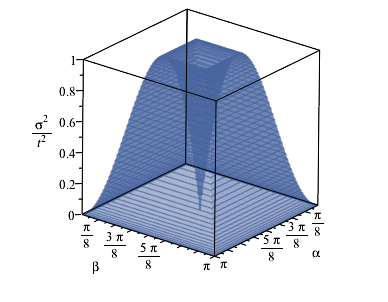}
\caption{Rescaled variance of the coinless quantum walk on the one-dimensional lattice. The variance depends only on the parameters $\alpha$ and $\beta$.} \label{fig:sigma}
\end{figure}

%For example, let $\alpha = -\frac{\pi}{2}$, $\beta = \frac{2\pi}{3}$ and $\phi_1 = \phi_2 = 0$. From the previous equations we obtain $\left< x\right>_t \sim t$ and $\left< x^2\right>_t \sim 2t^2$. Thus, the variance $\sigma^2\sim t^2$. 
The variance is
\begin{equation}
\sigma^2 = (2t-\left< x\right>_t)\left< x\right>_t,
\end{equation}
which simplifies asymptotically to
\begin{equation}
\frac{\sigma^2}{t^2} = \left\{\left.\begin{array}{rl} 4\cos\beta(1-\cos\beta), & \quad (\alpha,\beta) \in \textrm{I} \\-4\cos\beta(1+\cos\beta), & \quad (\alpha,\beta) \in \textrm{II} \\4\cos\alpha(1-\cos\alpha), &\quad (\alpha,\beta) \in \textrm{III} \\-4\cos\alpha(1+\cos\alpha). &\quad (\alpha,\beta) \in \textrm{IV}\end{array}\right. \right. \grafico 
\end{equation}
Notice that $\sigma^2$ does not depend on parameters $\phi_1$ and $\phi_2$. The maximum value of $\sigma^2$ is $1$, which is achieved at the points $(\alpha,\beta)$  either for specific values $\beta \in \left\{\frac{\pi}{3},\frac{2\pi}{3}\right\}$ and the interval  $\alpha \in \left[\frac{\pi}{3},\frac{2\pi}{3}\right]$ or for specific values  $\alpha \in \left\{\frac{\pi}{3},\frac{2\pi}{3}\right\}$ and the interval $\beta \in \left[\frac{\pi}{3},\frac{2\pi}{3}\right]$, as can be seen in Fig.~\ref{fig:sigma}.

The maximum asymptotic value of the mean is $\left< x\right>_t = 2t$, which is obtained when $\alpha = \beta = \frac{\pi}{2}$. In this case $\sigma^2 = 0$, which shows that the wave function does not spread moving ballistically rightward. The mean is zero only for $\alpha = 0$ or $\beta = 0$, which also produces localized walks.  The only way to obtain a non-trivial symmetric walk is by starting with a non-local initial condition such as $\left(\ket{0}+i\ket{1}\right)/\sqrt{2}$.

The results of this section can be compared with the moments of the coined model described in the Appendix.

\section{Two-dimensional lattice}\label{sec:2d}
From the tessellation for the two-dimensional lattice shown in Fig.~\ref{fig:tess}, we define the vectors
\begin{eqnarray}
\ket{u^0_{xy}} &=& \frac{1}{2}\sum_{x',y'=0}^{1}\ket{2x+x',2y+y'},\\
\ket{u^1_{xy}} &=& \frac{1}{2}\sum_{x',y'=0}^{1}\ket{2x+x'+1,2y+y'+1}.
\end{eqnarray}
The propagator is ${\cal{U}} = U_1U_0$, where
\begin{equation}
U_{0,1} = 2\sum_{x,y=-\infty}^{\infty}\ket{u_{xy}^{0,1}}\bra{u_{xy}^{0,1}}-\cal{I}.
\end{equation}
Ref.~\cite{Portugal:2014} describes the spectral decomposition of the evolution operator for the two-dimensional case. %Note that we are considering the infinite case. For the premises of the Fourier basis presented in Section~\ref{sec:fsol}, a few changes have been made in the eigenvalues and eigenvectors described by~\cite{Portugal:2014}. 
The Fourier basis in this case is
\begin{eqnarray}
\ket{\psi_{kl}^{0}} &=& \sum_{x,y = -\infty}^{\infty}e^{-i(2xk+2yl)}\ket{2x,2y},\\
\ket{\psi_{kl}^{1}} &=& \sum_{x,y = -\infty}^{\infty}e^{-i(2xk+(2y+1)l)}\ket{2x,2y+1},\\
\ket{\psi_{kl}^{2}} &=& \sum_{x,y = -\infty}^{\infty}e^{-i((2x+1)k+2yl)}\ket{2x+1,2y},\\
\ket{\psi_{kl}^{3}} &=& \sum_{x,y = -\infty}^{\infty}e^{-i((2x+1)k+(2y+1)l)}\ket{2x+1,2y+1}.
\end{eqnarray}
From the Fourier basis we can generate the $4\times 4$ reduced matrix
\begin{equation}
{{\cal{U}}_{kl}} = \left[\begin{array}{cccc} \frac{\cos{k}\cos{l}}{e^{-i(k+l)}} & \frac{\sin{k}\cos{l}}{ie^{-ik}} & \frac{\cos{k} \sin{l}}{ie^{-il}} & \sin{k}\sin{l} \vspace{2mm}\\\vspace{2mm}\frac{\sin{k}\cos{l}}{ie^{-ik}} & \frac{\cos{k}\cos{l}}{e^{i(k-l)}} & -\sin{k}\sin{l} & ie^{-il}\cos{k}\sin{l} \\\vspace{2mm}\frac{\cos{k}\sin{l}}{ie^{-il}} & -\sin{k}\sin{l} & \frac{ \cos{k}\cos{l}}{e^{i(k-l)}} & ie^{-ik}\sin{k} \cos{l} \\\sin{k}\sin{l} & ie^{-il}\cos{k} \sin{l} & ie^{-ik}\sin{k}\cos{l} & e^{-i{(k+l)}}\cos{k}\cos{l}\end{array}\right],
\end{equation}
whose eigenvalues are $1$ and $e^{\pm i\theta}$ where
\begin{equation}\label{eq:costheta}
\cos\theta = 2\cos^2k\cos^2l - 1.
\end{equation}
The eigenvectors associated with eigenvalue $1$ are
\begin{equation}
\ket{w^{0}_{kl}} = \frac{1}{2c^+}\left[\begin{array}{c}\sin(k-l) \\\sin l - \sin k \\\sin l -\sin k \\\sin(k-l)\end{array}\right], \ket{w^{1}_{kl}} = \frac{1}{2c^-}\left[\begin{array}{c}\sin(l-k) \\\sin l + \sin k \\-\sin l -\sin k \\\sin(k-l)\end{array}\right],
\end{equation}
where $(c^{\pm})^2 = (1\pm \cos k\cos l)(1\mp \cos(k-l))$. 
The eigenvectors associated with eigenvalue $e^{-i\theta}$ is
 \begin{equation}
\ket{w^{2}_{kl}} = \frac{1}{2c}\left[\begin{array}{c}-\epsilon\sqrt{c-\epsilon\sin k\cos l}\sqrt{c-\epsilon\cos k\sin l} \\\sqrt{c-\epsilon\sin k\cos l}\sqrt{c+\epsilon\cos k\sin l} \\\sqrt{c+\epsilon\sin k\cos l}\sqrt{c-\epsilon\cos k\sin l} \\\epsilon\sqrt{c+\epsilon\sin k\cos l}\sqrt{c+\epsilon\cos k\sin l}\end{array}\right],	
\end{equation}
where $c^2 = 1-\cos^2k\cos^2l$ and $\epsilon$ is the sign of $\cos k\cos l$. 
The eigenvectors associated with eigenvalue $e^{i\theta}$ are obtained by inverting the sign of $\epsilon$ in the eigenvectors associated with $e^{-i\theta}$.

Let us start the walk with the initial condition
\begin{equation}\label{eq:IC2D}
\ket{\psi(0)}=a\ket{00}+b\ket{01}+c\ket{10}+d\ket{11},
\end{equation}
which corresponds to a state in the blue cell that contains the origin,  see  Fig.~\ref{fig:tess}.
Using the expressions for eigenvectors $\ket{w^{\beta}_{kl}}$ and Eq.~(\ref{eq:costheta}), %and that
%\begin{equation}
%\epsilon\cos k\cos l = \sqrt{\cos^2k\cos^2l},
%\end{equation}
we obtain from Eq.~(\ref{eq:expikx_t}) that the first moments are  
\begin{eqnarray}\label{eq:oddm2}
\left< x\right>_t &=& D_2\left(\abs{a}^2+\abs{b}^2-\abs{c}^2-\abs{d}^2+a\overline{b}+b\overline{a}-d\overline{c}-\right.\left.c\overline{d}\right)t+O(1),\\
\left< y\right>_t &=&  D_2\left(\abs{a}^2-\abs{b}^2+\abs{c}^2-\abs{d}^2-b\overline{d}-d\overline{b}+c\overline{a}+\right.\left.a\overline{c}\right)t+O(1).      %\sim 0.7267604554t,
\end{eqnarray}
The second moments are
\begin{eqnarray}\label{eq:evenm2}
\left< x^2\right>_t &=& 2\left(D_2+\left(-\frac{3}{\pi}+1\right)\left(a\overline{c}+c\overline{a}+d\overline{b}+b\overline{d}\right)+\right.\nonumber\\
&&\left.\left(-\frac{7}{3\pi}+1\right)\left(a\overline{b}+b\overline{a}+d\overline{c}+c\overline{d}\right)+\right.\nonumber\\
&&\left.\left(\frac{10}{3\pi}-1\right)\left(b\overline{c}+c\overline{b}+a\overline{d}+d\overline{a}\right)\right)t^2+O(t),\\
\left< y^2\right>_t &=& 2\left(D_2+\left(-\frac{7}{3\pi}+1\right)\left(a\overline{c}+c\overline{a}+d\overline{b}+b\overline{d}\right)+\right.\nonumber\\
&&\left.\left(-\frac{3}{\pi}+1\right)\left(a\overline{b}+b\overline{a}+d\overline{c}+c\overline{d}\right)+\right.\nonumber\\
&&\left.\left(\frac{10}{3\pi}-1\right)\left(b\overline{c}+c\overline{b}+a\overline{d}+d\overline{a}\right)\right)t^2+O(t),
\end{eqnarray}
where
\begin{equation}\label{eq:D2}
    D_2=1-\frac{2}{\pi}.
\end{equation}
The maximum value of the coefficient of the total mean square displacement  $\sigma^2 = \sigma_x^2+\sigma_y^2$ is $8{D_2}\approx 2.91$, which is obtained for more than one value of parameters $a,b,c,d$. In the real case, there is only one assignment, which is $a=b=c=d=1/2$.

The results of this section can be compared with the moments of the coined model described in the Appendix.

\section{Conclusions}

Using the method of Fourier transforms and generating functions, we have obtained an analytical expression for the $n$th moment of the probability distribution of the coinless quantum walk model on $d$-dimensional lattices in terms of the $n$th derivative of the eigenvalues of the reduced propagator.  We have analyzed in details the mean square displacement for the one- and two-dimensional lattices. For the one-dimensional case we have taken a localized initial condition and analyzed the most generic coinless walk with a 2-site tessellation. The mean square displacement $\sigma^2$ depends only on parameters $\alpha$ and $\beta$ and the values of those parameters that produce the maximum $\sigma^2$ are depicted in Fig.~\ref{fig:sigma}. For the two-dimensional lattice we have taken non-localized initial conditions and analyzed the coinless walk with a 4-site tessellation using the simplest choice of basis vectors. 
The real assignment for the initial condition which gives the maximum mean square displacement is  the uniform one. 

\section*{Acknowledgements}
RAMS acknowledges financial support from Faperj E-45/2013. RP thanks Faperj (grant n.~E-26/102.350/2013) and CNPq (grant n.~304709/2011-5, 4741\-43/2013-9, and 400216/2014-0). SB acknowledges financial support from the
U. S. National Science Foundation through grant DMR-1207431.

\bibliographystyle{unsrt}       % APS-like style for physics
\bibliography{bib_moments2}   % name your BibTeX data base

\begin{thebibliography}{10}

\bibitem{Portugal:2013}
R.~Portugal.
\newblock {\em Quantum walks and search algorithms}.
\newblock Springer, New York, 2013.

\bibitem{Aharonov:1993}
Y.~Aharonov, L.~Davidovich, and N.~Zagury.
\newblock Quantum random walks.
\newblock {\em Physical Review A}, 48(2):1687--1690, 1993.

\bibitem{Patel:2005}
Apoorva Patel, K.~S. Raghunathan, and Pranaw Rungta.
\newblock Quantum random walks do not need a coin toss.
\newblock {\em Phys. Rev. A}, 71:032347, Mar 2005.

\bibitem{Ambainis:2001}
Andris Ambainis, Eric Bach, Ashwin Nayak, Ashvin Vishwanath, and John Watrous.
\newblock One-dimensional quantum walks.
\newblock In {\em Proceedings of the Thirty-third Annual ACM Symposium on
  Theory of Computing}, STOC '01, pages 37--49, New York, NY, USA, 2001. ACM.

\bibitem{Nayak:2000}
Ashwin Nayak and Ashvin Vishwanath.
\newblock Quantum walk on the line, 2000.
\newblock \url{arXiv:quant-ph/0010117v1}.

\bibitem{Konno:2002}
Norio Konno.
\newblock Quantum random walks in one dimension.
\newblock {\em Quantum Information Processing}, 1(5):345--354, October 2002.

\bibitem{SBJ14}
M.~\ifmmode \check{S}\else \v{S}\fi{}tefa\ifmmode~\check{n}\else
  \v{n}\fi{}\'ak, I.~Bezd\ifmmode~\check{e}\else \v{e}\fi{}kov\'a, and I.~Jex.
\newblock Limit distributions of three-state quantum walks: The role of coin
  eigenstates.
\newblock {\em Phys. Rev. A}, 90:012342, Jul 2014.

\bibitem{FB14}
Stefan Falkner and Stefan Boettcher.
\newblock Weak limit of the three-state quantum walk on the line.
\newblock {\em Phys. Rev. A}, 90:012307, Jul 2014.

\bibitem{WKKK08}
Kyohei Watabe, Naoki Kobayashi, Makoto Katori, and Norio Konno.
\newblock Limit distributions of two-dimensional quantum walks.
\newblock {\em Phys. Rev. A}, 77:062331, Jun 2008.

\bibitem{Ampadu:2011}
Clement Ampadu.
\newblock Limit theorems for quantum walks associated with hadamard matrices.
\newblock {\em Phys. Rev. A}, 84:012324, Jul 2011.

\bibitem{PBHMPBK14}
T.~J. Proctor, K.~E. Barr, B.~Hanson, S.~Martiel,
  V.~Pavlovi\ifmmode~\acute{c}\else \'{c}\fi{}, A.~Bullivant, and V.~M. Kendon.
\newblock Nonreversal and nonrepeating quantum walks.
\newblock {\em Phys. Rev. A}, 89:042332, Apr 2014.

\bibitem{Falk:2013}
M.~Falk.
\newblock Quantum search on the spatial grid, 2013.
\newblock \url{arXiv:quant-ph/1303.4127}.

\bibitem{Patel:2010}
Apoorva Patel, K.~S. Raghunathan, and Md.~Aminoor Rahaman.
\newblock Search on a hypercubic lattice using a quantum random walk. ii.
  $d=2$.
\newblock {\em Phys. Rev. A}, 82:032331, Sep 2010.

\bibitem{Ambainis:2013}
A.~Ambainis, R.~Portugal, and N.~Nahimov.
\newblock Spatial search on grids with minimum memory.
\newblock \url{arXiv:quant-ph/1312.0172}.

\bibitem{Portugal:2014}
R.~Portugal, S.~Boettcher, and S.~Falkner.
\newblock One-dimensional coinless quantum walks, 2014.
\newblock \url{arXiv:quant-ph/1408.5166v2}.

\bibitem{TFRK03}
Ben Tregenna, Will Flanagan, Rik Maile, and Viv Kendon.
\newblock Controlling discrete quantum walks: coins and initial states.
\newblock {\em New Journal of Physics}, 5(1):83, 2003.

\end{thebibliography}

%\begin{thebibliography}{}
% etc
%\end{thebibliography}

\section*{Appendix A}

\subsection*{\textbf{A.1 First and second moments of the Hadamard DTQW}}

The shift operator for the Hadamard DTQW is
\begin{equation}\label{inf_S}
    S= \sum_{x=-\infty}^\infty \ket{0}\bra{0}\otimes\ket{x+1}\bra{x}+\ket{1}\bra{1}\otimes\ket{x-1}\bra{x},
%%%S=\ket{0}\bra{0}\otimes \sum_{x=-\infty}^\infty \ket{x+1}\bra{x}+\ket{1}\bra{1}\otimes \sum_{x=-\infty}^\infty \ket{x-1}\bra{x}.
\end{equation}
where the coin operator is
\begin{equation}\label{inflat_pq_Hadamard}
    H = \frac{1}{\sqrt 2}\begin{bmatrix}
                           1 & 1 \\
                           1 & -1 \\
                         \end{bmatrix}
\end{equation}
and the initial state with generic coin state is 
$$\ket{\psi(0)}=\left(\cos\frac{\alpha}{2}\ket{0}+\textrm{e}^{i\phi}\sin\frac{\alpha}{2}\ket{1}\right)\,\ket{x=0}.$$

Using the same techniques employed in this work, we can calculate the expressions of the first and second moments for the coined model, which are
\begin{eqnarray}
\langle x \rangle &=& D_1\,\left(1+ \sin \alpha \cos \phi\right)\, t + O(1),\label{eq:1stmoment}\\
\langle x^2 \rangle &=& D_1 \, t^2 + O(t),
\end{eqnarray}
where
\begin{equation}\label{eq:D1}
    D_1=1 - \frac{1}{\sqrt 2}.
\end{equation}
Note that $D_1$ is a characteristic number of the Hadamard walk. The second moment does not depend on the parameters $\alpha,\phi$ of the initial condition and is characterized by $D_1$ asymptotically. On the other hand, the standard deviation $\sigma(t)=\sqrt{\langle x^2 \rangle-\langle x \rangle^2}$ depends on the parameters of the initial condition. It is not possible to obtain a sub-ballistic walk, since the smallest value of the coefficient of $t$ in the standard deviation is $\sqrt{D_1-2\,D_1^2}\approx 0.35$, which is obtained when $\alpha=\pi/2$, $\phi=0$. The largest coefficient is $\sqrt{D_1}\approx 0.54$ when $\alpha=\pi/2$, $\phi=\pi$.

\subsection*{\textbf{A.2 First and second moments of the two-dimensional Grover DTQW}}

The shift operator for the two-dimensional regular lattice is~\cite{Portugal:2013}
\begin{equation}\label{pq_S_x_y_0}
    S\ket{i,j}\ket{x,y}=\ket{i,j}\ket{x+(-1)^j(1-\delta_{i,j}),y+(-1)^j\delta_{i,j}}.
\end{equation}
The Grover coin is given by
\begin{eqnarray}
  G &=&
\frac{1}{2}\begin{bmatrix}
          -1 & \,\,\,\,1 & \,\,\,\,1 & \,\,\,\,1 \\
          \,\,\,\,1 & -1 & \,\,\,\,1 & \,\,\,\,1 \\
          \,\,\,\,1 & \,\,\,\,1 & -1 & \,\,\,\,1 \\
          \,\,\,\,1 & \,\,\,\,1 & \,\,\,\,1 & -1 \\
        \end{bmatrix}
\end{eqnarray}
and the initial state with generic coin state is 
$$\ket{\psi(0)}=\big(a\ket{00}+b\ket{01}+c\ket{10}+d\ket{11}\big)\,\ket{x=0,y=0}$$
with the constraint $|a|^2+|b|^2+|c|^2+|d|^2=1$.

Using the same techniques employed in this work, we can calculate the expressions of the first moments, which are
\begin{eqnarray}
\langle x \rangle &=& \frac{D_2}{2}\, \left(|b|^2-|c|^2- \Re\{(a+d)(\mybar{b}-\mybar{c})\}\right)\, t + O(1),\label{eq:1stmomentx}\\
\langle y \rangle &=& \frac{D_2}{2} \,  \left(|a|^2-|d|^2- \Re\{(a-d)(\mybar{b}+\mybar{c})\}\right)\, t + O(1),\label{eq:1stmomenty}
\end{eqnarray}
where
\begin{equation}\label{eq:D2_2}
    D_2=1-\frac{2}{\pi}.
\end{equation}
and the asymtotic expressions of the second moments, which are
\begin{eqnarray}
\frac{\langle x^2 \rangle}{t^2} &\simeq& {\frac {1+ \left| b \right|^{2}+
 \left| c \right|^{2}}{6\pi }}+{\frac { \left| 
a+d \right|^{2}}{12\pi }}+ \left( \frac{1}{2}-{\frac {19}{12\pi}}\right)  
\left| b-c \right|^{2}- \nonumber \\
&& \left(\frac{1}{2}-\frac{4}{3\pi} \right) \Re\left\{  \left( a+d \right) \left( \mybar{b}+\mybar{c} \right)  \right\}, \\
\frac{\langle y^2 \rangle}{t^2} &\simeq& {\frac {1+ \left| a \right|^{2}+
 \left| d \right|^{2}}{6\pi }}+{\frac { \left| 
b+c \right|^{2}}{12\pi }}+ \left( \frac{1}{2}-{\frac {19}{12\pi}}\right)  
\left| a-d \right|^{2}- \nonumber \\
&& \left(\frac{1}{2}-\frac{4}{3\pi} \right) \Re\left\{  \left( b+c \right) \left( \mybar{a}+\mybar{d} \right)  \right\},
\end{eqnarray}
where $\Re\{x\}$ is the real part of $x$. The maximum value of the coefficient of the standard deviation $\sigma=\sqrt{\sigma_x^2+\sigma_y^2}$ is $\sqrt{D_2}\approx 0.60$, which is obtained for more than one value of parameters $a,b,c,d$. In the real case, there is only one assignment, which is $a=-b=-c=d=1/2$. This result was obtained numerically in Ref.~\cite{TFRK03}. The minimum value of the coefficient of the standard deviation is $\sqrt{10/3\pi - 1}\approx 0.25$, which is obtained by setting  $a=b=c=d=1/2$.

\end{document}